\documentclass[a4paper, superscriptaddress, twocolumn, prl, showpacs, longbibliography]{revtex4-1}
\usepackage[utf8]{inputenc}
\usepackage{dcolumn}
\usepackage{bm}
\usepackage[colorlinks=true,citecolor=blue,linkcolor=blue,pdfhighlight =/O]{hyperref}
\usepackage{graphicx}
\usepackage{amsmath}
\usepackage{cancel}

\usepackage{dsfont}
\usepackage{mathtools}
\usepackage[dvipsnames]{xcolor}
\usepackage{soul}
\usepackage[normalem]{ulem}
\usepackage{bm}
\usepackage{amsmath,amssymb,amsfonts,latexsym,fancyhdr,graphicx}
\usepackage{simplewick}
\usepackage[english]{babel}
\usepackage{graphicx}
\usepackage{subfigure}
\usepackage{xcolor}
\usepackage{color}
\usepackage{verbatim}

\begin{document}

\title{Superconductivity of $\alpha-$gallium probed on the atomic scale by normal and Josephson tunneling}

\author{C. Fohn}
\affiliation{\mbox{Univ.} Grenoble Alpes, CNRS, Grenoble INP, Institut N\' eel, 38000 Grenoble, France}

\author{D. Wander}
\affiliation{\mbox{Univ.} Grenoble Alpes, CNRS, Grenoble INP, Institut N\' eel, 38000 Grenoble, France}

\author{D. Nikoli\'c}
\affiliation{Institut f\"ur Physik, Universit\"at Greifswald, Felix-Hausdorff-Strasse 6, 17489 Greifswald, Germany}
\affiliation{Fachbereich Physik, Universit\"at Konstanz, D-78457 Konstanz, Germany}

\author{S. Garaud\'ee}
\affiliation{\mbox{Univ.} Grenoble Alpes, CNRS, Grenoble INP, Institut N\' eel, 38000 Grenoble, France}

\author{H. Courtois}
\affiliation{\mbox{Univ.} Grenoble Alpes, CNRS, Grenoble INP, Institut N\' eel, 38000 Grenoble, France}

\author{W. Belzig}
\affiliation{Fachbereich Physik, Universit\"at Konstanz, D-78457 Konstanz, Germany}


\author{C. Chapelier}
\affiliation{\mbox{Univ.} Grenoble Alpes, CEA, Grenoble INP, IRIG/DEPHY/PHELIQS, 38000 Grenoble, France}

\author{V. Renard}
\affiliation{\mbox{Univ.} Grenoble Alpes, CEA, Grenoble INP, IRIG/DEPHY/PHELIQS, 38000 Grenoble, France}

\author{C. B. Winkelmann}
\affiliation{\mbox{Univ.} Grenoble Alpes, CNRS, Grenoble INP, Institut N\' eel, 38000 Grenoble, France}
\affiliation{\mbox{Univ.} Grenoble Alpes, CEA, Grenoble INP, IRIG/DEPHY/PHELIQS, 38000 Grenoble, France}

\begin{abstract}
We investigate superconducting gallium in its $\alpha$ phase using scanning tunneling microscopy and spectroscopy at temperatures down to about 100 mK.
High-resolution tunneling spectroscopies using both superconducting and normal tips show that superconducting $\alpha$-Ga
is accurately described by Bardeen-Cooper-Schrieffer theory, with a gap $\Delta_{\rm Ga}$ = 163 $\mu$eV on the $\alpha-$Ga(112) facet, with highly homogeneous spectra over the surface, including atomic defects and step edges. Using a superconducting Pb tip, we furthermore study the low-bias conductance features of the Josephson junction formed between tip and sample. The features are accurately described by dynamical Coulomb blockade theory, highlighting $\alpha-$Ga as a possible platform for surface science studies of mesoscopic superconductivity.

\end{abstract}
\maketitle

Atomically clean superconducting surfaces are attracting great interest in recent years as possible platforms for engineering complex quantum states, resulting for instance from the competition of superconductivity and quantum magnetism \cite{heinrich2018single}. 
To this end, a growing number of superconducting substrates has been investigated by low-temperature scanning probe techniques. In particular, the combination of scanning tunneling microscopy and spectroscopy (STM/STS) allows mapping the orbital properties of quantum states at well defined energies \cite{ruby2016orbital,choi2017mapping}. 
Furthermore, the ability to manipulate and arrange individual atoms {\em in situ} into chains and two-dimensional structures with the STM tip allows for the emergence of a whole new field in condensed matter dedicated to quantum engineering the hybridization of multiple states and thereby creating tunable energy bands \cite{nadj2014observation,ruby2015end,schneider2021topological,schneider2023proximity}.

Superconducting gallium has not been investigated to date by low-temperature STM/STS, due to its low critical temperature $T_c=1.083$ K \cite{Ga_1964phillips_SCheatcap, GA_1966gregory1966_STS}, combined with a melting point near room temperature, adding difficulties to the in vacuo surface preparation. Yet, its multiple metastable crystalline phases make it an intriguing material that could have promising applications in plasmonics \cite{gutierrez2019plasmonic}. Furthermore, the coexistence of covalent and metallic interatomic bonds make gallium for instance the elemental conductor with the highest spatial anisotropy of electrical conductance \cite{defrain1977etats}.  Recently, the observation of a very large magnetoresistance in $\alpha$-Ga at low temperatures led to claims that it might have topological properties \cite{Ga_2018chen_magnetoresistance}. 

In this work, we study the superconducting properties of the $\alpha$ phase of gallium by STM/STS down to temperatures of about 100 mK. We find excellent agreement with Bardeen-Cooper-Schrieffer (BCS) theory, with a gap $\Delta_{\rm Ga}$ = 163 $\mu$eV on the (112) facet. 
The superconducting properties such as the shape and the magnitude of the gap are very robust over the scanned area, including atomic defects and step edges,  which rules out the claim 
of topological superconductivity \cite{Ga_2018chen_magnetoresistance}. Using a superconducting Pb tip and reducing the tunnel resistances, we investigate the low-energy current-voltage characteristics of the Ga-Pb Josephson junction formed between tip and sample. The transport features are well described by the model of voltage-biased ultra-small Josephson junctions, in the frame of dynamical Coulomb blockade theory.

The low-temperature crystallographic ground state of Ga is the orthorhombic $\alpha$ phase, where other phases ($\beta$, $\gamma$, $\delta$, amorphous,...) can be obtained in thin films \cite{cohen1967strong} or by supercooling or -heating at the solid-liquid transition of gallium \cite{feder1966hysteresis}.   
The $\alpha$ phase is formed of covalently bonded Ga$_2$ dimers, so that $\alpha$-Ga is considered a molecular solid, presenting a pseudo-gap and thus a small density of states at the Fermi level \cite{Ga_1991gong_molecularmetallicstructure,Ga_zuger1992atomic}. 
This leads to a small $T_c$ in $\alpha$-Ga, while the other main phases display superconductivity at larger temperatures, around 6 K or above \cite{feder1966hysteresis}. Early tunneling experiments on $\alpha-$Ga crystals using macroscopic contacts revealed a superconducting gap $\Delta_{\rm Ga}\approx160\,\mu$eV, varying by $\pm5\%$ depending on the crystal orientation \cite{yoshihiro1969GaBCSSpectroscopy,yoshihiro1970GaBCSSpectroscopy}, an observation which was explained using Fermi-surface selection rules for tunneling \cite{gregory1971GaBCSSpectroscopy}.
Furthermore, the structural properties of the various gallium phases and their surface orientations have only scarcely been investigated by scanning probe techniques, and mostly near ambient temperature \cite{Ga_zuger1992atomic, Ga_zuger1992stm,Ga_gruetter1997Ga112} with a main focus on the low-index facets of $\alpha$-Ga, such as the (100), (010), (111), and (112) orientations.

\begin{figure*}
    \centering
    \includegraphics[width=1.9\columnwidth]{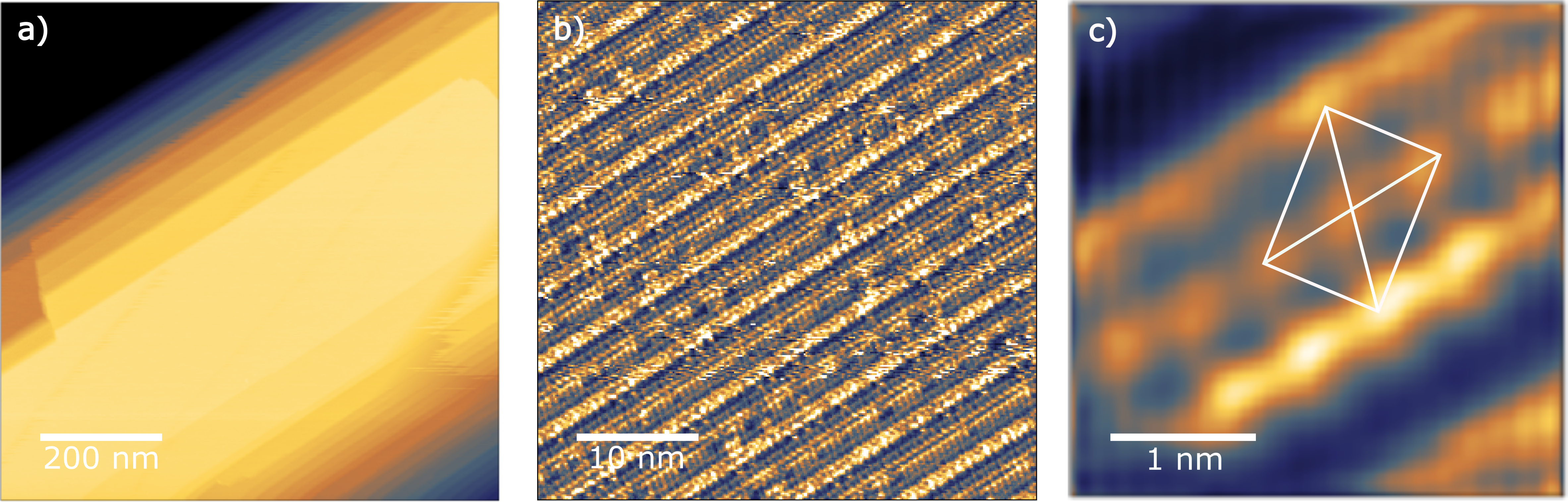}
    \caption{ (a) Constant-current STM topograph displaying atomic terraces on the $\alpha$-Ga surface ($I=20$ pA, $V_b=800$ mV). (b)  Constant-height image of the tunneling current, revealing the striped surface reconstruction on a terrace characteristic of the (112) orientation ($V_b=3$ mV). (c) Constant-height close-up on the atomic structure of the stripes. The unit cell (rhombus) is tilted with respect to the stripes. Images taken at $T=4.3$ K.}
    \label{fig:images}
\end{figure*}

The experimental setup is based on an inverted dilution refrigerator, named {\it sionludi}, operating at temperatures down to 50 mK, and coupled to an ultra-high vacuum (UHV) surface preparation chamber, with base pressure $2\times10^{-10}$ mbar. The vacuum chamber hosting the cryostat and the STM has a few non-conflat vacuum flanges, limiting the operating pressure to about $3\times10^{-9}$ mbar. Except for the STM head (Tribus Ultra from Scienta Omicron), the setup is entirely homemade. A calibrated resistive thermometer (Cernox sensor) located directly on the STM indicates a temperature reading that is systematically a few tens of mK above that of the cryostat temperature (determined by a ruthenium-oxyde sensor) and saturates at 115 mK, which is also about the lower operational limit of Cernox temperature sensors. Because the STM is suspended on springs and thermalized to the cryostat via a mesh of silver wires, we use in the remainder the STM thermometer reading rather than the mixing chamber temperature as the experimentally relevant temperature scale $T$. 

For STM/STS, we use a PtIr tip, which we clean by field emission on a Au(111) sample, and test on a Pb(111) crystal at 4.2 K. Superconducting Pb tips are obtained by deeply indenting the PtIr tip into the same substrate. After indentation, proper tip conditions are verified in tunneling spectroscopies on the Pb(111) surface, which must yield a total spectroscopic gap of width $2(\Delta_{\rm sample}+\Delta_{\rm tip})/e=4\Delta_{\rm Pb}/e\approx 5.3\pm0.1$ meV. This criterion, together with a Pb-Pb differential conductance at the gap edge exceeding by at least 10 times the high-bias tunneling conductance, indicates a good superconducting tip. Eventually, field emission on the Au(111) crystal allows recovering a normal conducting tip.

\begin{figure*}
    \centering
    \includegraphics[width=2.05\columnwidth]{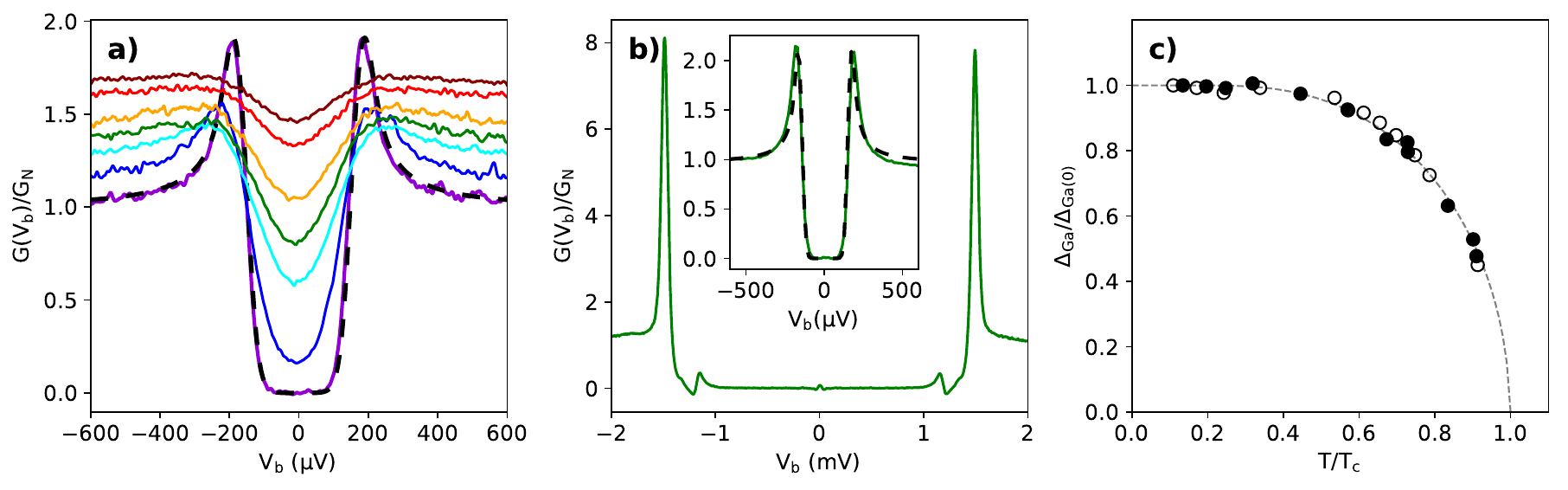}
    \caption{ \textbf{(a)} Tunneling spectroscopies (vertically shifted for clarity) of the superconducting gap of Ga using a normal PtIr tip, at various cryostat temperatures varying from 50 to 970 mK ($G_N = 625$ nS). The dashed line is a fit of the lowest temperature dataset with BCS theory, yielding $\Delta_{\rm Ga}(0)=163\,\mu$eV and $T_{\rm noise}=80$ mK (see text). \textbf{(b)} Tunneling conductance between the Ga surface and a Pb-coated tip at $T=617$ mK ($G_N=1.1\,\mu$S, $V_{ac} = 15\, \mu$V). A weak zero-bias Josephson peak as well as thermally activated subgap conductance peaks at $V_b=\pm(\Delta_{\rm Pb}$ - $\Delta_{\rm Ga})/e$ are observed. The inset shows the Ga density of states obtained by deconvolution, yielding a gap of 151 $\mu$eV at this temperature. The BCS fit (grey dashed) was obtained assuming an electrical noise level as found from $T_{\rm noise}$ in (a). \textbf{(c)} Temperature dependence of $\Delta_{\rm Ga}$: Comparison of the gap determined from NIS (bullets) and SIS' (circles) spectroscopies, with $\Delta_{\rm Ga}(0)=163\,\mu$eV and $T_c=1.08$ K. The dashed line is the BCS prediction.}
    \label{fig:spectroscopy}
\end{figure*}

Available Ga single crystals were difficult to fit into the experimental sample space. Thus, the Ga sample under investigation was obtained by gently melting and slowly re-solidifying a 99.9999 \% (6N) pure Ga ingot on a CuBe STM sample plate. In the UHV chamber, the surface is cleaned by ion milling with Ne at a pressure in the $10^{-5}$ mbar range for 30 minutes, at ambient temperature, that is, 5 to 10 degrees below the Ga melting temperature $T_m=29.8\, ^{\circ}$C. Given the proximity to the melting point, the ion milling power must be kept low and no further annealing is required for restructuring the surface. The sample is then transferred {\it in vacuo} to the STM maintained at liquid helium temperature, and further cooled to the dilution refrigerator's base temperature within less than two hours.

Because the sample is not a single crystal, both the crystal phase and the surface orientation are a priori unknown. From the spectroscopic properties, further discussed below, it is nevertheless clear that the sample is indeed in the $\alpha$ phase. Furthermore, the observed topographic structures coincide closely with those reported for the Ga(112) surface \cite{Ga_gruetter1997Ga112}. 
The surface topographic images display atomic terraces (Fig. \ref{fig:images}a), with step heights appearing as integer multiples of $h=2.0\pm0.1$ \AA, to be compared with the tabulated crystallographic value of 1.956 \AA, and $2.2\pm0.2$ reported in prior STM experiments \cite{Ga_gruetter1997Ga112}. On the atomic terraces, which can be up to several 100 nm wide, a striped surface reconstruction with an alternation of two bright rows with different intensities is observed, with a transverse period of $37.6\pm0.8$ \AA \, (Fig. \ref{fig:images}b). This distinctive feature was also observed in studies at higher temperatures on Ga(112), both using LEED and STM, with a period of $37\pm 1$ \AA \,\cite{Ga_gruetter1997Ga112}.
The rows resemble knitwear due to Ga$_2$-molecules lying at higher levels, which were explained by a missing-row model  \cite{Ga_gruetter1997Ga112}. This striped structure is due to a characteristic $(4\times1)$ surface reconstruction of the (112) facet, consisting of three rows which are separated by two deeper lying ones. 
On the atomic scale, the unit cell is observed to be tilted with respect to the stripes (Fig. \ref{fig:images}c), which are in the $\left[1\bar 10\right]$ direction, in good agreement with the crystallographic structure.

We  perform tunneling spectroscopies of the superconducting $\alpha$-Ga surface using the lock-in technique to measure the differential junction conductance $G(V_b)=dI/dV_b$, which we normalize to its value $G_N$ at large bias voltage $V_b$. With the pristine PtIr tip on superconducting Ga, we are thus in the situation of a normal metal-insulator-superconductor (NIS) tunnel junction, whereas with the superconducting tip an SIS'-type junction connects two distinct superconductors. 
The tunneling spectroscopies obtained using the normal tip reveal the superconducting density of states in Ga and can be nicely fitted at all temperatures using Fermi's golden rule and assuming a density of states in $\alpha$-Ga described by the BCS theory (Fig. \ref{fig:spectroscopy}a). We find a gap $\Delta_{\rm Ga}(0)=163 \,\mu$eV in the low-temperature limit, without recurring to any broadening parameter except for an effective temperature of the tip $T_{\rm eff}=\sqrt{T^2+ T_{\rm noise}^2}$. We find that a $T_{\rm noise}=80$ mK, imputable to residual electronic noise, provides excellent agreement between our experimental temperature scale and the observed spectroscopic broadening at all temperatures. Consequently, $T_{\rm eff}$ reaches about 140 mK at the STM base temperature of 115 mK. 

 The experimental temperature dependence of $\Delta_{\rm Ga}(T)$ is shown in Fig. \ref{fig:spectroscopy}c, and follows very closely the theoretical prediction from BCS theory. The gap can be extrapolated to close at $1.08\pm0.05$ K, where the main uncertainty stems from the thermometer accuracy, from which we deduce eventually $2\Delta_{\rm Ga}(0)/k_BT_c=3.50\pm0.3$, with $k_B$ the Boltzmann constant. For completeness, we have investigated the spatial dependence of the superconducting gap over the $\alpha$-Ga surface. We did not observe any measurable variations here, not even close to structural defects or atomic terrace edges, indicating a homogeneous superconducting state. The combination of the above spectroscopic observations on the atomic scale  provides strong  evidence that the $\alpha$ phase of Ga shows no deviations from weak-coupling BCS superconductivity, in contrast with speculations based on magnetoresistance measurements \cite{Ga_2018chen_magnetoresistance}.

\begin{figure}[t]
    \centering
    \includegraphics[width=1.0\columnwidth]{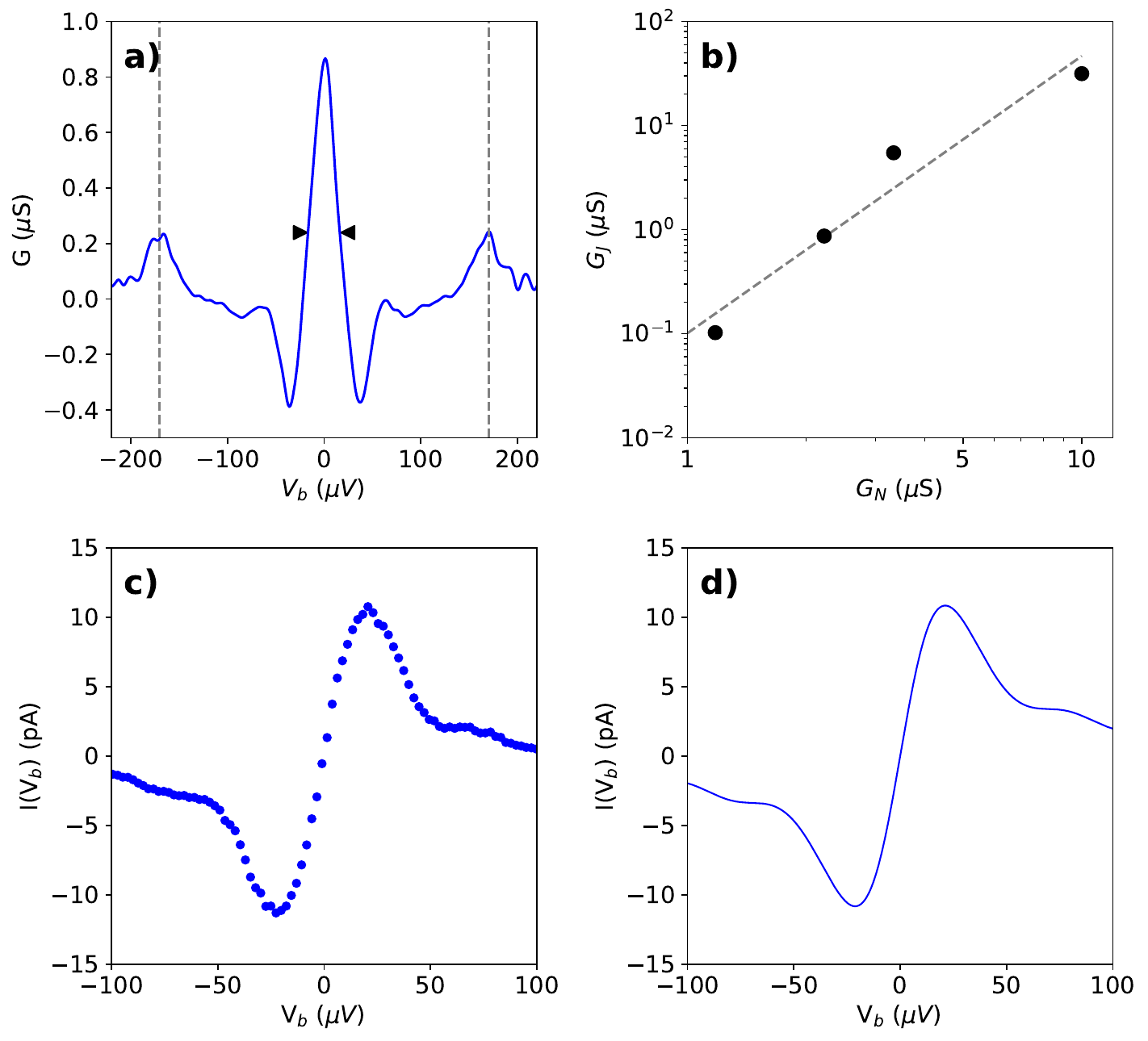}
    \caption{ (a) Small-bias tunneling spectroscopy on the $\alpha-$Ga surface taken with a superconducting Pb tip, with $G_N=2.2\,\mu$S and at $T=120$ mK, displaying the characteristic low-bias Josephson behavior. The conductance peaks at $\pm \Delta_{\rm Ga}/e$ (dashed vertical lines) correspond to the resonant pair tunneling between the Pb superconducting condensate and the $\alpha-$Ga coherence peaks. The full width at half maximum (black arrows) is 31 $\mu$V. (b) Josephson peak height for different normal state conductances. The dashed line is the expected trend $\propto G_N^2$ (see text). (c) Experimental $I(V_b)$ curve at $T=120$ mK, and (d) corresponding theory plot (see text). Note that the vertical scale of the calculation matches very well the experimental data without the use of any rescaling factors.
}
    \label{fig:Josephson}
\end{figure}


As a prerequisite for the Josephson experiments, and because tunneling spectroscopies of a superconductor using a normal tip display a significant thermal broadening at temperatures near $T_c$, we have repeated the experiment using a superconducting Pb tip. We measure the Pb tip density of states by tunneling spectroscopies on Au(111) at very low temperature, from which we determine $\Delta_{\rm tip}(0)=1.34\pm 0.015$ meV, close to the 1.35 meV bulk value. In the temperature range of interest for studying superconductivity in $\alpha$-Ga, $\Delta_{\rm tip}$ can be considered constant. 
When moving to the Ga sample, even at intermediate temperatures the SIS' tunneling current between Pb and Ga displays very sharp differential conductance peaks at $V_b=\pm (\Delta_{\rm tip}+\Delta_{\rm Ga})/e$ (Fig. \ref{fig:spectroscopy}b), where $e$ is the elementary charge. Furthermore, at temperatures not significantly smaller than the $T_c$ of Ga, thermally activated additional peaks at $V_b=\pm (\Delta_{\rm tip}-\Delta_{\rm Ga})/e$ are also visible (Fig. \ref{fig:spectroscopy}b). These features are completely absent at lower temperatures. Here again, the tunneling characteristics can be very well fitted using a Fermi golden rule transport integral between two superconductor. This allows deconvoluting the data and retrieving the Ga density of states even near its $T_c$ with very high resolution, as seen in the inset. Note that because the $\sim 15\,\mu$eV uncertainty on our initial determination of $\Delta_{\rm tip}$ (see above) can lead to a large relative error on $\Delta_{\rm Ga}$, we impose $\Delta_{\rm Ga}(0)$ to be the same as found from the NIS spectroscopies, which leads in our case to refining $\Delta_{\rm tip}$ by a $10 \,\mu$eV downshift. The extracted temperature dependence of $\Delta_{\rm Ga}$ is plotted in Fig. \ref{fig:spectroscopy}c, and superimposes perfectly with the results obtained using the normal conducting tip.

We now turn to the investigation of the Josephson coupling between a superconducting Pb tip and the $\alpha$-Ga surface. Indeed, for normal state tunneling conductances $G_N$ in the $\mu$S range or above, second order processes between two superconductors can allow for a finite Josephson coupling across the junction. This leads to a zero-bias conductance peak in SIS' tunnel junctions (with same or different gaps), which is already visible in Fig. \ref{fig:spectroscopy}b and studied in greater detail and for a larger $G_N$ in Fig. \ref{fig:Josephson}a. 
Furthermore, the peaks visible near $\pm \Delta_{\rm Ga}/e$ correspond to the resonant tunneling of electron pairs between the Ga coherence peaks and the Pb superconducting condensate, that is, the lowest-order multiple Andreev reflection process of an SIS' junction with two different gaps \cite{kuhlmann1994andreev}.

The theoretical description of the low-bias transport in voltage-biased Josephson junctions calls for the picture of inelastic Cooper-pair tunneling. In this picture, Josephson and charging effects enter in competition, with respective characteristic energies $E_J=\hbar I_c/(2e)$, with $I_c$ the critical current, and $E_C=2e^2/C$, with $C$ the capacity of the tunnel junction. For ultra-small tunnel junctions, such that $E_J<E_C$, this picture leads to the dynamical Coulomb blockade effect, also called $P(E)$ theory \cite{schon1990,Ingold94}. It was successfully applied to STM Josephson junctions \cite{Randeria2016, jaeck2016, esat2023determining}. 
The presence of an electromagnetic (EM) environment coupled to the junction leads to the fluctuations of the superconducting phase causing the charge tunneling. The resulting Cooper-pair-mediated current reads~\cite{Averin1990, IngoldNazarov}
\begin{equation}
	I_s(V) = \frac{\pi e E_J^2}{\hbar} \left[\mathcal{P}(2eV)-\mathcal{P}(-2eV)\right],
 \label{eqn:Is}
\end{equation}
where $\mathcal{P}(E)$ is the probability for a Cooper pair to emit a photon of energy $E$ into the EM environment during the inelastic tunneling process across the junction.
Furthermore, the critical current of the Ga-Pb tunnel junction is given by the Ambegaokar-Baratoff formula with two different gaps~\cite{Ambegaokar1963}, 
\begin{equation}
 \begin{split}
	I_c(T) =  &\frac{2\pi k_B T G_N}{e}
 \\  &\times \sum_{n \in{\mathbb N}}\frac{\Delta_{\rm Ga}(T)}{\sqrt{\omega_n^2+\Delta_{\rm Ga}(T)^2}}\frac{\Delta_{\rm tip}}{\sqrt{\omega_n^2+\Delta_{\rm tip}^2}},
 	\label{eqn:Ic}
 \end{split}
\end{equation}
where the $\omega_n=(2n+1)\pi k_BT$ are the positive fermionic Matsubara frequencies. 
From the combination of Eqs. (\ref{eqn:Is}) and (\ref{eqn:Ic}), it is expected that the zero-bias conductance associated to the Josephson effect 
$G_J= (dI_s/dV_b)|_{V_b=0}$
must be $\propto E_J^2\propto G_N^2$. Experimentally, this is indeed well obeyed at base temperature (Fig. \ref{fig:Josephson}b), within experimental scatter due to instabilities of the tip. As discussed in more detail in the Appendix, we calculate the $\mathcal{P}(E)$ function by modelling the junction environment as a damped $LC$ transmission line at a temperature $T_{\rm env}=T$, which allows describing in full detail the Josephson current-voltage characteristics (Fig. \ref{fig:Josephson}c,d). In particular, the transmission lines' modes lead to the smooth bump in the $I(V_b)$ characteristics visible near $\pm70\,\mu$V, confirming the accuracy of this picture. 

 In conclusion, we have presented an STM/STS study of superconductivity of the $\alpha$-Ga(112) surface down to millikelvin temperatures. Despite its low $T_c$, the superconducting gap and its temperature dependence are characterized with high precision and found to be very homogeneous and BCS like. Beyond its single-particle excitation spectrum, we probe the superconducting condensate by quasiparticle- and Josephson tunneling from a superconducting tip. Both the quasiparticle tunneling spectra and their temperature dependence, as well as the Josephson data are well described by theory. Finally, and except for its rather large topographic corrugation, due to the striped $(4\times1)$ surface reconstruction, the $\alpha-$Ga(112) surface is a novel and promising platform for further investigations of nanoscale superconductivity.

 \section{Acknowledgements}

We thank Katharina Franke for discussions and G\'erard Lapertot for providing the gallium. This work was funded by the joint ANR-DFG grant JOSPEC (ANR-17-CE30-0030) and by the European Union under the Marie Sklodowska-Curie grant agreement no. 766025 (QuESTech). D.N. and W.B. acknowledge support from the EU’s Horizon 2020 research and innovation program under Grant Agreement No. 964398 (SUPERGATE) and from Deutsche Forschungsgemeinschaft (DFG; German Research Foundation) via SFB1432 (Project No. 425217212).

 \section{Appendix: Modelling the electromagnetic environment}

 Several quantum circuit experiments have allowed solidly establishing the $P(E)$ theory by careful on-chip engineering of the environmental impedance $Z(\omega)$ and its temperature $T_{\rm env}$ \cite{steinbach2001direct}. In STM experiments the situation is more complicated because the environmental impedance and even its temperature are not as well controlled. A few Josephson STM experiments operating at low temperatures around or below 100 mK obtained very good agreement with $P(E)$ theory by assuming either a damped $LC$-type electromagnetic environment in thermal equilibrium with the STM junction \cite{Randeria2016, jaeck2016}, or an ohmic environment at higher temperatures \cite{esat2023determining}.

 To start with, $\mathcal{P}(E)$ depends crucially on the junction’s EM environment, both through its frequency-dependent impedance $Z(\omega)$, as well as its temperature $T_\mathrm{env}$, which can be different from that of the junction, $T$. 
 As it has been pointed out in Ref.~\cite{Jaeck2015}, Cooper pair tunneling can be mediated by two parallel channels: (i) fluctuation due to the EM environment itself and (ii) thermal voltage fluctuation across the junction of capacitance $C$ given by $\sqrt{\bar{v}^2}=\sqrt{k_BT_\mathrm{env}/C}$. This leads to writing $\mathcal{P}(E)$ as the convolution of two functions, $P(E )$ and $P^\prime (E)$, accounting for both above sources of fluctuations, respectively:

\begin{equation}
	\label{eqn:P(E)_full}
	\mathcal{P}(E) = \int_{-\infty}^{\infty} dE' P(E)P'(E-E').
\end{equation}

Here, the second of the functions above involves the charging energy $E_C=2e^2/C$ and is simply given by a Gaussian shape

\begin{equation}
	P'(E)=\sqrt{\frac{1}{4\pi E_Ck_BT_\mathrm{env}}} e^{-E^2/4E_Ck_BT_\mathrm{env}}.
\end{equation}
In the above expression appears a characteristic energy scale $\sqrt{4\pi E_Ck_BT_\mathrm{env}}$, resulting from the geometric average of the thermal and the charging energies. In other words, the knowledge of $P^\prime(E)$ does not allow determining independently $T_\mathrm{env}$ and $E_C$ (that is, $C$). 

 The expression for $P(E)$ function describing the fluctuations of the phase caused by the environment is given by
\begin{equation}
	\label{eqn:P(E)}
	P(E)=\frac{1}{2\pi\hbar}\int_{-\infty}^{\infty}dt \exp\left[4J(t)+\frac{i}{\hbar}Et\right],
\end{equation}
where $J(t)=\langle[{\varphi}(t)-{\varphi}(0)]{\varphi}(0)\rangle$
is the equilibrium correlation function of the phase $\varphi(t)=(e/\hbar)\int_{-\infty}^t V(t')dt'$ with $V(t)$ being the voltage across the junction. The $J(t)$ function depends on the total impedance of the system seen by the junction, $Z_t(\omega)$, and reads
\begin{align}
	\label{eqn:J(t)}
	J(t)=\, &2\int_0^\infty \frac{d\omega}{\omega}\frac{\Re[Z_t(\omega)]}{R_K}\\&\times
	\bigg\{\coth\bigg(\frac{\hbar\omega}{2k_{\rm B}T_\mathrm{env}}\bigg)[\cos(\omega t)-1]-i\sin(\omega t)\bigg\}\nonumber,
\end{align}
where $R_K=h/e^2\approx 25.8$ k$\Omega$ is the von Klitzing constant. The total impedance $Z_t(\omega)$ is given by
\begin{equation}
	\label{eqn:Z_t}
	Z_t(\omega)=\frac{1}{i\omega C + Z^{-1}(\omega)}, 
\end{equation}
where $C$ is the junction's capacitance and $Z(\omega)$ is the environmental impedance discussed below.

 Since there is no universal recipe for describing the STM tip as an EM environment, we follow previous reports on the topic~\cite{Jaeck2015,Randeria2016} and model it as a finite $LC$ transmission line. The corresponding total impedance $Z_t(\omega)$ [see Eq.~\eqref{eqn:Z_t}] has the form~\cite{Ingold94}

\begin{equation}
	\label{eqn:Z_line}
	\frac{Z_t(\omega)}{R_K}=\rho \frac{1+\frac{i}{r}\tan(\frac{\pi}{2}\nu)}{\left[1-\kappa\frac{\omega}{\omega_0}\tan(\frac{\pi}{2}\nu)\right]+ir\left[\kappa \nu+\tan(\frac{\pi}{2}\nu)\right]}.
\end{equation}
Here, $\rho=R_L/R_K$, $r$ is a dimensionless quality factor, $\nu=\omega/\omega_0$ with $\omega_0$ denoting the $n=0$ eigenfrequency of the line, and $\kappa=\omega_0Z_0C$ is the parameter related to the capacitive shunt of the junction, with $Z_0\approx377~\Omega$ the vacuum impedance. For reproducing the experimental data we choose the following values of the above parameters: $\rho=0.008$, $r=0.55$, $\omega_0/E_C=1.5$, and $\kappa=0.27$. Moreover, for the charging energy of the junction we take $E_C\approx 0.2 \Delta_{\rm Ga}(0) \approx 35~\mu$eV and we assume that the environmental temperature is in equilibrium with the STM, at $T_\mathrm{env}=120~$mK. These assumptions, which are also close to the parameters used in Refs. \cite{jaeck2016,Randeria2016}, allow eventually obtaining the very good agreement with the experimental data shown in Figs.~\ref{fig:Josephson}c,d.

  \bibliography{mybib.bib}
 
\end{document}